\documentclass[%
 reprint,
 amsmath,amssymb,
 aps,
 prx,
 superscriptaddress,
 longbibliography,
]{revtex4-2}

\usepackage{graphicx}
\usepackage{dcolumn}
\usepackage{bm}

\usepackage{amsmath}
\newcommand*{\rttensor}[1]{\overline{\overline{#1}}}

\begin{document}

\preprint{APS/123-QED}

\title{Artificial Neural Network with Physical Dynamic Metasurface Layer \\for Optimal Sensing}

\author{Philipp del Hougne}
 \email{philipp.delhougne@gmail.com}
 \affiliation{Institut de Physique de Nice, CNRS UMR 7010, Universit\'{e} C\^{o}te d'Azur, Nice, France}
\author{Mohammadreza F. Imani}%
\affiliation{Department of Electrical and Computer Engineering, Duke University, Durham, North Carolina, USA}%
\author{Aaron V. Diebold}%
\affiliation{Department of Electrical and Computer Engineering, Duke University, Durham, North Carolina, USA}%
\author{Roarke Horstmeyer}
\affiliation{Biomedical Engineering Department, Duke University, Durham, North Carolina, USA}%
\author{David R. Smith}
\affiliation{Department of Electrical and Computer Engineering, Duke University, Durham, North Carolina, USA}%

\date{\today}

\begin{abstract}

We address the fundamental question of how to optimally probe a scene with electromagnetic (EM) radiation to yield a maximum amount of information relevant to a particular task.
Machine learning (ML) techniques have emerged as powerful tools to extract task-relevant information from a wide variety of EM measurements, ranging from optics to the microwave domain. 
However, given the ability to actively illuminate a particular scene with a programmable EM wavefront, it is often not clear what wavefronts optimally encode information for the task at hand (e.g., object detection, classification). 
Here, we show that by integrating a physical model of scene illumination and detection into a ML pipeline, we can jointly learn optimal sampling and measurement processing strategies for a given task. 
We consider in simulation the example of classifying objects using microwave radiation produced by dynamic metasurfaces. 
By integrating an analytical forward model describing the metamaterial elements as coupled dipoles into the ML pipeline, 
we jointly train analog model weights with digital neural network weights.
The learned non-intuitive illumination settings yield  a higher classification accuracy using fewer measurements. 
On the practical level, these results are highly relevant to emerging context-aware systems such as autonomous vehicles, touchless human-interactive devices or within smart health care, where strict time constraints place severe limits on measurement strategies. 
On the conceptual level, our work serves as a bridge between wavefront shaping and tunable metasurface design on the physical layer and ML techniques on the processing layer.

\end{abstract}

\maketitle

\section{Introduction}

Wave-based sensing is of fundamental importance in countless applications, ranging from medical imaging to non-destructive testing \cite{sebbah2001waves}. Currently, it is emerging as key enabling technology for futuristic ``context-aware'' concepts like autonomous vehicles \cite{hasch2012millimeter,khalid2017integrated}, ambient-assisted living facilities \cite{rashidi2013survey,adib2015smart} 
and touchless human-computer 
interaction (HCI) devices \cite{molchanov2015multi,google_touchless_interaction}. In these context-aware settings, an important goal is often to achieve the highest possible accuracy for a given task, such as recognizing a hand gesture, with as few measurements as possible. 
Minimizing the number of measurements can help improve a wide number of metrics - for example, speed, power consumption, and device complexity. It is also crucial in a variety of specific contexts -  for instance, to limit radiation exposure (e.g., x-ray imaging \cite{xray}), to adhere to strict timing constraints caused by radiation coherence or unknown movements in a biological context \cite{liu2015optical,wang2015focusing,blochet2017focusing}, or to make real-time decisions in automotive security \cite{katrakazas2015real,khalid2017integrated}. 

In all of the above applications, ``active" illumination is sent out from 
the 
device to interact with the scene of interest before 
the reflected waves are captured by the sensor. 
The resulting measurements are then processed to achieve a particular goal. Usually, acquisition and processing are treated and optimized separately. For instance, the spatial resolution of a LIDAR system on an autonomous vehicle is often optimized to be as high as possible, while its resulting measurements are subsequently processed to detect pedestrians with as high an accuracy as possible. Recently, machine learning (ML) techniques have dramatically improved the accuracy of measurements post-processing for complex tasks (like object recognition) without requiring explicit analytical instructions \cite{krizhevsky2012imagenet,satat2017object,sinha2017lensless,rivenson2017deep,borhani2018learning,li2018deep}. 

However, to date, the physical acquisition layers of context-aware systems have yet to reap the benefits of new ML techniques. At the same time, by separately optimizing acquisition hardware and post-processing software, most sensing setups are 
not tailored to their specific sensing task. Instead, as with the LIDAR example noted above, hardware is typically optimized to obtain a high-fidelity visual image for human consumption, thereby often ignoring available knowledge that could help to highlight information that is critical for ML-based analysis. 

Here, we address both of the above shortcomings with a new ``learned sensing'' 
paradigm for context-aware systems that allows for joint optimization of acquisition hardware and post-processing software. The result is a device that acquires non-imaging data that is optimized for a particular ML task. We consider the challenge of identifying settings of a reconfigurable metamaterial-based device emitting microwave patterns that can encode as much relevant information about a scene for subsequent ML-based classification with as few measurements as possible. However, as we will detail, this framework is general, flexible, and can impact a number of future measurement scenarios.

\section{Illumination Strategies in Wave-Based Sensing}

A number of prior works have attempted to optimize active illumination in the microwave, terahertz, and optical regimes to improve the performance of certain sensing tasks. The simplest approach in terms of the transceiver hardware is often to use random illumination, for instance, by leveraging the natural mode diversity available in wave-chaotic or multiply-scattering systems \cite{montaldo2005building,chan2008single,hunt2013metamaterial,liutkus2014imaging}. Random illuminations have a finite overlap that reduces the amount of information that can be extracted per additional measurement. A truly orthogonal illumination basis, such as the Hadamard basis \cite{fenn2000development,swift1976hadamard,watts2014terahertz,satat2017lensless}, has also been frequently used to overcome this (minor) issue, often at the cost of more complicated hardware.

These forms of ``generic" illumination often fail to efficiently highlight salient features for task-specific sensing, which is desirable to reduce the number of required measurements. In other words, they do not discriminate between relevant and irrelevant information for the task at hand. Task-specific optimal illumination can be challenging to determine, due to hardware constraints (e.g., few-bit wavefront modulation), possible coupling effects between different transceivers, and in particular a lack of insight into the inner workings of the ML network (i.e., the artificial neural network, ANN) used to process the acquired data for each task. So far, most attempts at task-specific tailored illuminations seek to synthesize illumination wavefronts matching the expected principal components of a scene \cite{jolliffe2011principal,neifeld2003feature,pca_proposal,li2019machine}. To outperform generic illumination, such an approach requires a sufficiently large aperture with a sufficient amount of tunable elements to synthesize wavefronts in reasonable agreement with the expected principal components. Moreover, a sufficient number of measurements has to be taken to cover the most important principal scene components.

Approaches based on a principal component analysis (PCA) of the scene can thus be interpreted as a step towards optimal wave-based sensing (see Fig.~S3 in the Supplemental Material); they work well under favorable conditions (large aperture, many tunable elements, unrestricted number of measurements). However, the fundamental challenge of extracting as much task-relevant information as possible using a general wave-based sensor thus remains open. Besides its fundamental interest, the question is also of high relevance to many practical applications: for instance, in automotive RADAR and LIDAR, the aperture size, the number of tunable illumination elements, and the measurement sampling rate over space and time are all highly restricted. 
In these constrained scenarios, we hypothesize that wave-based sensing can benefit from \textit{joint} optimization of data acquisition and processing.

Inspired by recent works in the optical domain \cite{chakrabarti,horstmeyer2017convolutional}, this work interprets data acquisition as a trainable physical layer that we integrate directly into an ANN pipeline. By training the ANN with a standard supervised learning procedure, we can simultaneously determine optimal illumination settings to encode relevant scene information, along with a matched post-processing algorithm to extract this information from each measurement --- automatically taking into account any constraints on transceiver tuning, coupling and the number of allowed measurements. 

As noted above, we apply our concept to classification tasks with microwave sensors. These tasks are a crucial stepping stone towards numerous  context-aware systems e.g. in smart homes, for hand gesture recognition with HCI devices, for concealed-threat identification in security screening, and in autonomous vehicles \cite{hasch2012millimeter,khalid2017integrated,rashidi2013survey,adib2015smart,molchanov2015multi,google_touchless_interaction}. Microwave frequencies can operate through optically opaque materials such as clothing, are not impacted by external visible lighting and scene color, minimally infringe upon privacy (unlike visual cameras) and may eventually help sense through fog, smoke and ``around-the-corner'' \cite{google_touchless_interaction}.

We focus on optimally configuring dynamic metasurface hardware, a promising alternative to more traditional antenna arrays for beam-forming and wavefront shaping \cite{fenn2000development}. Dynamic metasurfaces are electrically large structures patterned with metamaterial elements that couple the modes of an attached waveguide or cavity to the far field  \cite{sleasman2015dynamic,imani2018two}. Reconfigurability is achieved by individually shifting each metamaterial element's resonance frequency, for instance, with a PIN diode \cite{TimCELC}. Compared to a traditional antenna array that uses amplifiers and phase shifters, the inherent analog multiplexing makes dynamic metasurface hardware much simpler, less costly and easier to integrate in many applications.

To demonstrate our proposed Learned Integrated Sensing Pipeline (LISP)
, we jointly optimize the illumination and detection properties of dynamic metasurface transceivers, along with a simple neural network classifier, for the task of scene classification. In this work, we consider the dummy task of classifying ``handwritten'' metallic digits in simulation. Replacing this dummy task with a more realistic scenario, such as concealed-threat detection, hand-gesture recognition or fall detection, is conceptually straight-forward. 
To construct the LISP, we first formulate an analytical forward model which is possible due to the intrinsic sub-wavelength nature of the metamaterial elements. 
Second, we allow certain key parameters within the ``physical" forward model to act as unknown weights (here the reconfigurable resonance frequency of each metamaterial element), that we aim to optimize over. Third, we then merge this weighted physical model into an ANN classifier, and use supervised learning to jointly train the unknown weights in both to maximize the system's classification accuracy. Despite coupling between metamaterial elements and a binary tuning constraint, which would otherwise be challenging to account for in a standard inverse model, our ANN implicitly accounts for this during training and identifies simultaneously optimal illumination settings and processing structures to extract as much task-relevant information as possible. We are also easily able to use this model to compare the performance of our scheme with the aformentioned benchmarks of 
orthogonal and PCA-based illumination.

Interestingly, our LISP 
can also be interpreted in light of recent efforts to meet the exploding computational demands of ANNs with wave-based analog computing, which seeks to perform desired operations with waves as they interact with carefully tailored systems \cite{silva2014performing,zhu2017plasmonic,shen2017deep,del2018leveraging,estakhri2019inverse,hughes2019wave,zangeneh2019topological,ising_conti}. 
Our wave-based sensing scheme is essentially a hybrid analog-digital ANN in which the interaction of learned optimal wavefronts with the scene acts as a first processing layer. 
As we will show, data acquisition 
fulfills two processing tasks in our pipeline, being (i) trainable and (ii) highly compressive.

\section{Operation Principle}\label{SectionOperationPrinciple}

This section is outlined as follows: first, we introduce the hardware of our reconfigurable metasurface apertures. Second, we establish an analytical forward model of our sensor's physical layer 
built upon a compact description of the metamaterial elements as dipoles. It consists of three steps: (i) extracting each metamaterial element's dipole moment while taking tuning state and inter-element coupling into account; (ii) propagating the radiated field to the scene; (iii) evaluating 
the scattered field. Third, we outline the sensing protocol. Finally, fourth, we integrate analog (physical) and digital (processing) layers into a unique ANN pipeline and discuss how it can be trained and account for binary reconfigurability constraints. 

Note that this section seeks to give a clear and thorough overview of the physical layer and its integration into the ANN pipeline by only providing key equations; details, derivations and equations defining all variables are included in Section A of the Supplemental Material for completeness, following Refs.~\cite{laura2018,guy2015}.

\begin{figure}[t]
	\begin{center}
\includegraphics [width=\columnwidth] {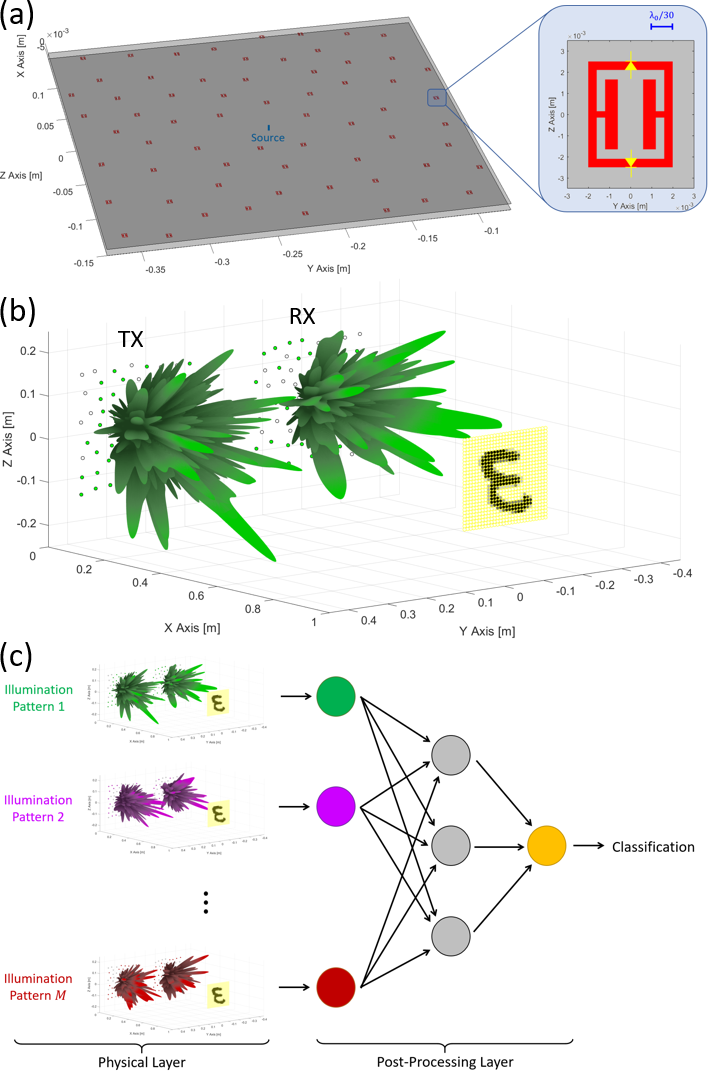}
	\caption{Schematic overview of operation principle. (a) \textbf{Dynamic metasurface.} $N=64$ tunable metamaterial elements are patterned into the upper surface of a planar waveguide. The inset shows the geometry of an example cELC metamaterial element that could be used in combination with PIN diodes (location and orientation indicated in yellow) to reconfigure the element. The waveguide is excited by the indicated line source. (b) \textbf{Sensing setup.} The scene consists of a metallic digit in free space that is illuminated by a TX metasurface and the reflected waves are captured by a second RX metasurface. (c) \textbf{Sensing protocol.} The scene is illuminated with $M$ distinct TX-RX metasurface configurations, yielding a $1 \times M$ complex-valued measurement vector that is processed by an artificial neural network consisting of fully connected layers. The output is a classification of the scene.  }
	\label{fig_schematic}
	\end{center}
\end{figure}

\subsection{Dynamic Metasurface Aperture}

The reconfigurable metasurface that we consider for the generation of shaped wavefronts is depicted in Fig.~\ref{fig_schematic}(a). It consists of a planar metallic waveguide that is excited by a line source  (a coaxial connector in a practical implementation). $N$ metamaterial elements are patterned into one of the waveguide surfaces to couple the energy to free space. 
An example of a metamaterial element that could be used is the tunable complimentary electric-LC (cELC) element \cite{TimCELC} shown in the inset of Fig.~\ref{fig_schematic}(a). A possible tuning mechanism to individually configure each metamaterial element's radiation properties involves diodes. Then, the 
cELC element is resonant or not (equivalently, radiating or not radiating) at the working frequency of $f_0 = 10\ \mathrm{GHz}$ depending on the bias voltage of two PIN diodes connected across its capacitive gaps. 
The $N$ metamaterial elements are randomly distributed within a chosen aperture size ($30\ \mathrm{cm} \times 30 \ \mathrm{cm}$) but a minimum distance between elements of one free-space wavelength is imposed  \footnote{Separating the metamaterial elements by at least one free-space wavelengths avoids evanescent coupling between them.}.

Similar dynamic metasurfaces have previously been used to generate random scene illuminations for computational imaging \cite{sleasman2015dynamic,imani2018two}. Here, we will take full advantage of the ability to individually control each radiating element to purposefully shape the scene illuminations. 
The individual addressablility of each metamaterial element distinguishes these dynamic metasurfaces from other designs in the literature that simultaneously reconfigure all elements to redirect a beam \cite{guzman2012electronic,fu2017electrically,chen2019electronically}.
Our LISP could of course also be implemented with other wavefront shaping setups such as arrays of reconfigurable antennas \cite{fenn2000development}, reflect-arrays illuminated by a separate feed (e.g. a horn antenna) \cite{huang2008reflectarray,sievenpiper2002tunable,cui2014coding} or leaky-wave antennas with individually controllable radiating elements \cite{Lim2004,Gregoire2014}. Indeed, if our concept was transposed to the acoustic or optical regime, one would probably use an array of acoustic transceivers or a spatial light modulator, respectively. In the microwave domain, however, antenna arrays are costly and the use of reflect-arrays yields bulky setups. In contrast, dynamic metasurface hardware as the one in Fig.~\ref{fig_schematic}(a) benefits from its inherent analog multiplexing and is moreover compact, planar and can be fabricated using standard PCB processes. Note that although we focus for concreteness on an implementation that we consider advantageous, our ideas are not limited to this specific hardware.

\subsection{Compact Analytical Forward Model}

The formulation of a compact analytical forward model is enabled by the intrinsic sub-wavelength nature of the radiating metamaterial elements which enables a convenient description in terms of (coupled) dipole moments \cite{draine1994discrete,scher2009extracting,bowen2012using,landy2014two}. Ultimately, we will directly link a given metasurface configuration (specifying which elements are radiating) to the radiated field. Note that the possibility to analytically do so is a key advantage of this metasurface hardware over alternative devices with reconfigurable radiation pattern such as leaky reconfigurable wave-chaotic three-dimensional cavities \cite{dupre2015wave,del2016spatiotemporal,sleasman2016microwave}. For the latter, one could learn forward models based on near-field scans of radiated fields \cite{peurifoy2018nanophotonic,liu2018training}; yet the number of required equivalent source dipoles \cite{guy2015} (even with a coarse half-wavelength sampling of the aperture) is much higher than in our case where it is simply $N$.

\subsubsection{Dipole Moments}

Generally speaking, a point-like scatterer's magnetic dipole moment $\mathbf{m}$ is related to the incident magnetic field $\mathbf{H^{loc}}$ via the scatterer's polarizability tensor $\rttensor{\alpha}$:

\begin{equation}\label{eq1}
  \mathbf{m} (\mathbf{r}) = \rttensor{\alpha} \ \mathbf{H^{loc}} (\mathbf{r}) ,%
\end{equation}\label{dipole_polariz_field}

\noindent where $\mathbf{r}$ denotes the scatterer's location. In our case, using surface equivalent principles, an effective polarizability tensor of the metamaterial element embedded in the waveguide structure can be extracted \cite{lauraprb}. For the metamaterial element we consider, only the $\alpha_{yy}$ component is significant. 
In the following, we use $\alpha_{yy} = (1.5-3.5\mathrm{i}) \times 10^{-7}\ \mathrm{m}^3$ which is a typical polarizability value for a tunable cELC element \cite{lauraprb} \footnote{We work with identical metamaterial elements here but note that the formalism can deal with different polarizabilities for different elements, see also Section~\ref{sec_robustness}.}.
The tuning state of a given metamaterial element can be encoded in its polarizability: if the element is ``off'', its polarizability (at the working frequency) is zero and hence it does not radiate any energy. We thus use $t_{j,i}\alpha_{yy}$ as effective polarizability, where $t_{j,i} \in \{ 0, 1 \}$ is the tuning state of the $j$th metamaterial element in the $i$th metasurface configuration.

The local magnetic field $\mathbf{H^{loc}}$ that excites the metamaterial element is a superposition of the feed wave and the fields scattered off the other metamaterial elements. $\mathbf{H^{loc}}$ in Eq.~\ref{eq1} is thus itself a function of the metamaterial elements' magnetic dipole moments, yielding the following system of coupled equations \cite{laura2018}:

\begin{subequations}\label{HlocLHS}
\label{eq:whole}
\begin{eqnarray}
H^{loc}_y (\mathbf{r_j}) =  (t_{j,i}\alpha_{yy})^{-1} \ m_y(\mathbf{r_j}) ,\label{eq2a}
\end{eqnarray}
\begin{equation}
H^{loc}_y (\mathbf{r_j}) = H^{feed}_y (\mathbf{r_j}) + \sum_{j\neq k} {  G_{yy}(\mathbf{r_j},\mathbf{r_k}) \ m_y(\mathbf{r_k}) } ,\label{eq2b}
\end{equation}
\end{subequations}

\noindent where Eq.~\ref{eq2a} is a rearranged version of Eq.~\ref{eq1} and the index $i$ identifies the $i$th metasurface configuration. Equation~\ref{eq2b} expresses the local field at location $\mathbf{r_j}$ as sum of the feed wave $H^{feed}_y$ at $\mathbf{r_j}$ and the individual contributions from each of the other metamaterial elements at $\mathbf{r_k}\neq \mathbf{r_j}$ via the Green's functions $G_{yy}(\mathbf{r_j},\mathbf{r_k})$. Explicit expressions for $H^{feed}_y(\mathbf{r_j})$ and $G_{yy}(\mathbf{r_j},\mathbf{r_k})$ are provided in the Supplemental Material. To solve Eq.~\ref{HlocLHS} for the magnetic dipole moments we rewrite it in matrix form and perform a matrix inversion, as shown in Step 1A in the top line of Fig.~\ref{fig_principal}:

\begin{equation}\label{solution_for_M}
  \mathbf{M_i} = \{ \mathbf{A_i} - \mathbf{G} \}^{-1} \mathbf{F}   , %
\end{equation}

\noindent where $\mathbf{M_i} = \left[m_{y,i}(\mathbf{r_1}),m_{y,i}(\mathbf{r_2}),\dots,m_{y,i}(\mathbf{r_N}) \right]$, $\mathbf{A_i} = \mathrm{diag}\left( \left[t_{1,i}\alpha_{yy})^{-1},(t_{2,i}\alpha_{yy})^{-1},\dots, (t_{N,i}\alpha_{yy})^{-1}\right]\right)$
and $\mathbf{F} = \left[H^{feed}_y(\mathbf{r_1}),H^{feed}_y(\mathbf{r_2}),\dots,H^{feed}_y(\mathbf{r_N}) \right]$. 
The off-diagonal entry $(j,k)$ of $\mathbf{G}$ is $G_{yy}(\mathbf{r_j},\mathbf{r_k})$, and the diagonal of $\mathbf{G}$ is zero since the self-interaction terms are incorporated into the effective polarizabilities in $\mathbf{A_i}$ \footnote{A detailed discussion on effective polarizability and the self-interaction term of the Green's function can be found in Section~II of Ref.~\cite{lauraprb}.}.

Due to the metamaterial element interactions via the $G_{yy}(\mathbf{r_j},\mathbf{r_k})$ term, the mapping from tuning state to dipole moments is not linear. This is visualised with an example in Fig.~S1. 
Thus, ultimately the mapping from tuning state to radiated field cannot be linear, which is a substantial complication for most beam synthesis approaches; as we will see in Section~\ref{pipeline_section}, this does not pose any additional complication in our LISP.

\subsubsection{Propagation to Scene}

Having found a description of the metamaterial elements in terms of dipole moments, we can now go on to identify the wavefront impacting the scene for a given metasurface configuration. The sensing setup we consider is depicted in Fig.~\ref{fig_schematic}(b). 
A transmit (TX) metasurface like the one discussed above illuminates 
the scene. 
The scene, in our case, consists of a planar metallic digit of size $40 \ \mathrm{cm} \times 40 \ \mathrm{cm}$ in free space that is to be recognized at a distance of $1\ \mathrm{m}$. 

To compute the $i$th incident TX wavefront (corresponding to the $i$th metasurface configuration) at a location $\mathbf{\zeta}$ in the scene, we superimpose the fields radiated by each of the $N$ metamaterial element dipoles \cite{jackson,guy2015}: 

\begin{equation}\label{e_tx}
   \mathbf{E_i^{TX}}(\mathbf{\zeta}) = \frac{-\mathrm{i}\omega \mu_0}{4\pi} \sum_{j=1}^N \left( (\mathbf{m_i}(\mathbf{r_j}) \times \hat{\mathbf{\rho_j}} ) \left( \frac{-\mathrm{i}k}{\Gamma_j} - \frac{1}{\Gamma_j^2} \right) \mathrm{e}^{-\mathrm{i}k\Gamma_j} \right), %
\end{equation}

\noindent where $\omega=2\pi f_0$, $\Gamma_j = |\mathbf{\zeta} - \mathbf{r_j}|$, $\hat{\mathbf{\rho_j}}$ is a unit vector parallel to $\mathbf{\zeta}-\mathbf{r_j}$ and $\mathbf{m_i}(\mathbf{r_j})$ is the magnetic dipole moment of the $j$th metamaterial element in the $i$th metasurface configuration. Equation~\ref{e_tx} is the second step of our analytical forward model, as shown in Step 1B in Fig.~\ref{fig_principal}. 
Since the magnetic dipole moments only have a significant $y$-component, the radiated electric field's dominant component is along $z$.

\subsubsection{Measurement}

To complete the physical layer description, we need to identify the portion of TX fields that is reflected off the scene
and collected by the second receiving (RX) metasurface, as shown in Fig.~\ref{fig_schematic}(b). Since our scene is flat and reflections are primarily specular, the first Born approximation is a suitable description: the field reflected at location $\mathbf{\zeta}$ in the scene is $\mathbf{E}^{\mathrm{TX}}_i(\mathbf{\zeta})\ \sigma(\mathbf{\zeta})$, where $\sigma$ is the scene reflectivity.  The RX metasurface captures specific wave forms depending on its configuration; their shape is essentially defined via a time-reversed version of Eq.~\ref{e_tx}. The complex-valued signal measured for the $i$th pair of TX-RX configurations is thus \cite{guy2015}

\begin{equation}\label{measurement}
   g_i \propto \int_{scene}  \mathbf{E}^{\mathrm{TX}}_i(\mathbf{\zeta})  \cdot  \mathbf{E}^{\mathrm{RX}}_i(\mathbf{\zeta}) \ \sigma(\mathbf{\zeta}) \ \mathrm{d}\mathbf{\zeta}. %
\end{equation}

This is the third and final step of our analytical forward model, shown as Step 1C in Fig.~\ref{fig_principal}. Note that the scene $\sigma(\mathbf{\zeta})$ is ultimately sampled by $ \mathbf{\mathcal{I}}_i(\mathbf{\zeta}) = \mathbf{E}^{\mathrm{TX}}_i(\mathbf{\zeta})  \cdot  \mathbf{E}^{\mathrm{RX}}_i(\mathbf{\zeta})$; when loosely referring to ``scene illumination'' in this work, we mean this product of $\mathbf{E}^{\mathrm{TX}}_i(\mathbf{\zeta})$ and $ \mathbf{E}^{\mathrm{RX}}_i(\mathbf{\zeta})$. 
\footnote{Each measurement corresponds to a distinct pair of a TX and an RX pattern. This is different from previous computational imaging schemes with random metasurface configurations \cite{imani2018two} that defined a set of random configurations and took a measurement for each possible combination of TX and RX configurations within the set.} 
In practice, to compute the integral, we discretize the scene at the Rayleigh limit, 
that is the scene consists of a $28 \times 28$ grid of points with half-wavelength spacing. Each point's reflectivity value $\sigma(\mathbf{\zeta})$ is a gray-scale real value determined by the corresponding handwritten digit's reflectivity map. 

\subsection{Sensing Protocol}

Having outlined the physical layer of our sensing setup, we next consider the sensing protocol. A single measurement, depending on various factors such as the sensing task's complexity, the type of scene illumination but also the signal-to-noise ratio, may not carry enough information to successfully complete the desired sensing task \cite{del2018precise}. Hence, in general, we illuminate the scene with $M$ distinct patterns. Each pattern corresponds to a specific pair of TX and RX metasurface configurations. Since our scheme is monochromatic, each measurement yields a single complex value $g_i$. 

As shown in Fig.~\ref{fig_schematic}(c), our $1 \times M$ complex-valued measurement vector 
is fed into a processing ANN. The latter consists of two fully connected layers. Real and imaginary parts of the measurement vector are stacked and the resultant real-valued vector is the input to the first layer consisting of $256$ neurons with ReLu activation. This is followed by a second fully connected layer made up of $10$ neurons with SoftMax activation, yielding a normalized probability distribution as output. (See Supplemental Material for details.) The highest value therein corresponds to the classification result (one digit between ``0'' and ``9''). These are the two digital layers shown in Fig.~\ref{fig_principal}. This architecture was chosen without much optimization and still performs quite well; its performance was observed to not significantly depend on the chosen parameters, such as the number of neurons.

\subsection{Hybrid Analog-Digital ANN Pipeline}\label{pipeline_section}

\begin{figure*}
	\begin{center}
\includegraphics [width=18cm] {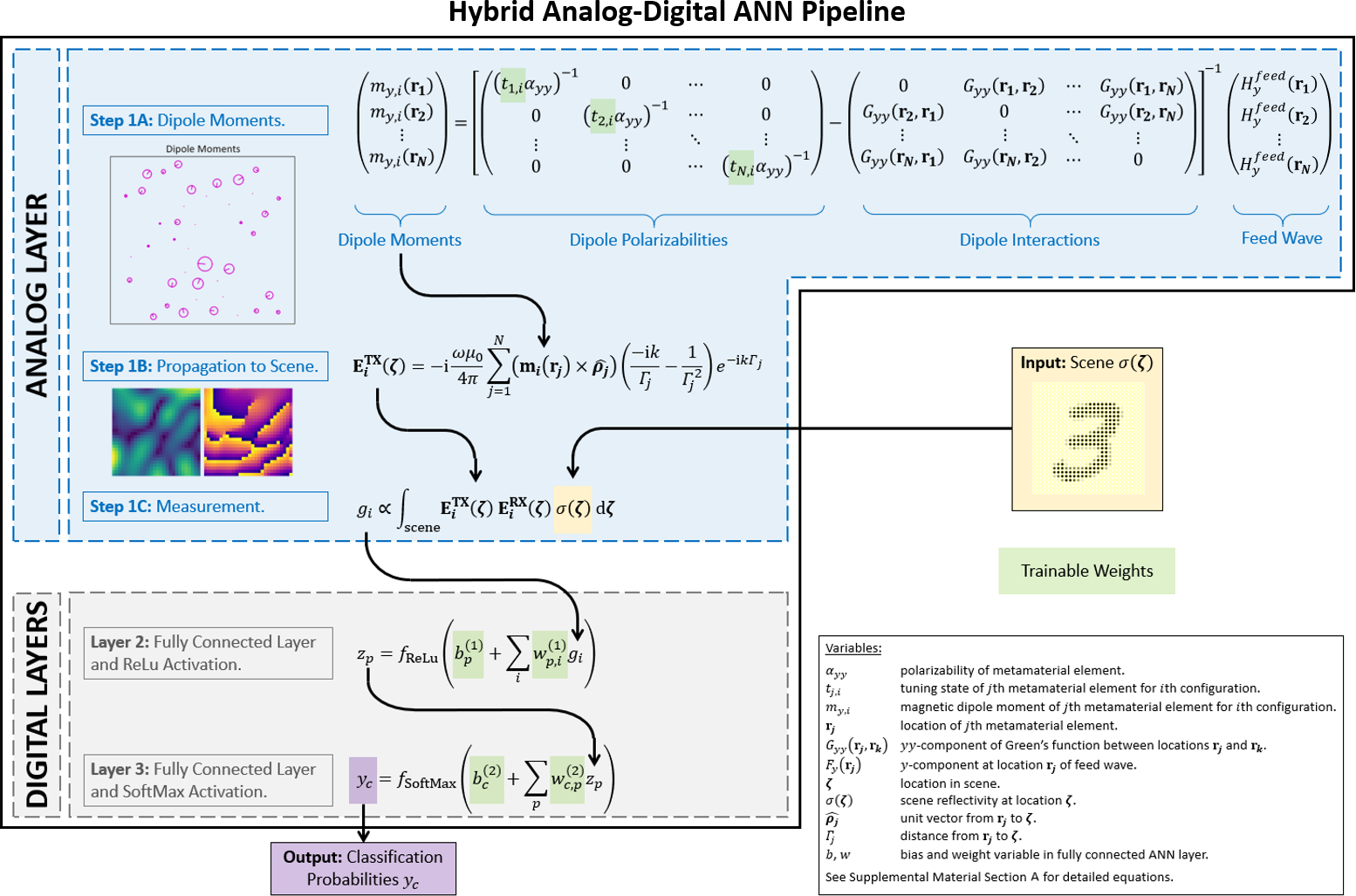}
	\caption{Overview of our Learned Integrated Sensing Pipeline (LISP). 
	The analog (physical) layer corresponds to the sensing setup introduced in Fig.~\ref{fig_schematic}(b). A scene is illuminated with a dynamic metasurface, and the reflected waves are captured with a second metasurface. The analytical forward model for the analog layer consists of three steps. First, each metamaterial element's magnetic dipole moment is calculated for a given metasurface configuration. The inset shows an example of calculated dipole moments which are represented as phasors, with the radius of the circle being proportional to their amplitude, and the line segment showing their phase. The circles are centered on the physical location of each metamaterial element. Second, the field radiated by these dipoles to the scene is computed. The inset shows amplitude (left) and phase (right) of a sample field illuminating the scene. 
	Third, the measurement is evaluated. Note that the figure contains the equations for Steps 1A and 1B only for the TX metasurface, for the sake of clarity; the RX equations are analogous. The measurement vector, consisting of complex-valued entries corresponding to different configurations of the TX-RX metasurfaces, is then processed by two fully connected layers consisting of $256$ and $10$ neurons, respectively. Finally, a classification of the scene is obtained as output. 
	Trainable weights in our hybrid analog-digital ANN pipeline are both in the analog and the digital layers and highlighted in green. During training, these are jointly optimized via error back-propagation. }
	\label{fig_principal}
	\end{center}
\end{figure*}

We are now in a position to assemble our pipeline consisting of an analog and two digital layers as outlined above (Fig.~\ref{fig_principal}). The input, a scene, is injected into the analog layer which contains trainable weights and is moreover highly compressive. The output from the analog layer, the measurement vector, continues to be processed by the digital layers which contain trainable weights as well. The final digital layer's output is the classification of the scene. By jointly training the analog and the digital weights, we identify illumination settings that optimally match the constraints and processing layer. 
Importantly, this means that the ANN will find an optimal compromise also in cases where the aperture size is small and few tunable elements are available, meaning that PCA modes cannot be synthesized accurately, and when the number of measurements is very limited, meaning that not all significant PCA modes can be probed.

While the digital weights ($b_p^{(1)}$, $w_{p,i}^{(1)}$, $b_c^{(2)}$ and $w_{c,p}^{(2)}$ in Fig.~\ref{fig_principal}) are real-valued variables drawn from a continuous distribution, our metasurface hardware requires the physical weights $t_{j,i}$ to be binary. At first sight, this constraint is incompatible with variable training by back-propagating errors through the ANN which relies on computing gradients \cite{backpropNature}. An elegant solution consists in the use of a temperature parameter that supervises the training, gradually driving the physical weights from a continuous to a discrete binary distribution \cite{chakrabarti}. The detailed implementation thereof is outlined in Section B.2 of the Supplemental Material. Note that we ultimately only have to formulate an analytical forward model; the fact that this model contains coupling effects and binary weight constraints does not bring about any further complications in our scheme.
The weights are trained using $60,000$ sample scenes from the reference MNIST dataset, as detailed in Section B.1 of the Supplemental Material.

In order to compare our LISP with the benchmarks of orthogonal and PCA-based illuminations, we solve the corresponding inverse design problems by only taking the analog layer of our pipeline and defining a cost function based on the scene illuminations (rather than the classification accuracy). The procedure is detailed in Section B.3 of the Supplemental Material.

\section{Results}

In this section, we analyze 
the sensing performance of our LISP 
and compare 
it to the three discussed benchmarks based on random, orthogonal and PCA-based illuminations. We  consider dynamic metasurfaces with $N=64$ or $N=16$ metamaterial elements and analyze whether the obtained optimal illuminations can be related to orthogonality or PCA-based arguments. 
Finally, we investigate the robustness to fabrication inaccuracies. 

\subsection{Sensing Performance}

\begin{figure*}
	\begin{center}
\includegraphics [width=18cm] {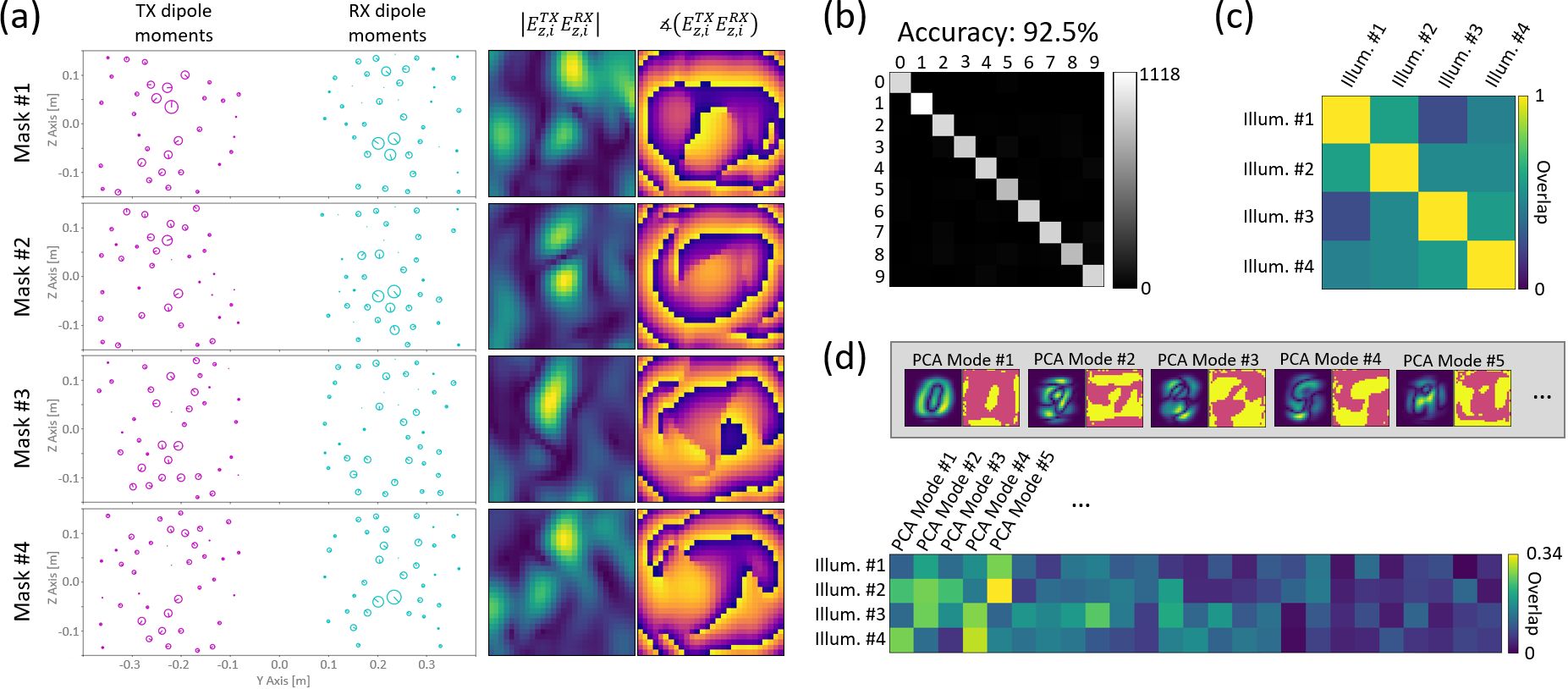}
	\caption{Analysis of 
	LISP illumination patterns for a single realization with $M=4$. 
	(a) For each of the four masks, the corresponding TX and RX dipole moments, and the magnitude and phase of the corresponding scene illuminations $E_{z,i}^{TX}E_{z,i}^{RX}$, $i\in\{1,2,3,4\}$, are shown. The dipole moment representations are as in the inset of Fig.~\ref{fig_principal}. The magnitude maps are normalized individually, the phase maps have a colorscale from $-\pi$ to $\pi$.
	(b) Confusion matrix evaluated on an unseen test dataset of $10,000$ samples. This realization achieved $92.5\%$ classification accuracy. 
	(c) Mutual overlap of the four scene illuminations. The average over the off-diagonal entries of the overlap matrix is $0.45$.
	(d) Overlap of the four scene illuminations with the first $25$ PCA modes. Note that the colorscale's maximum is $0.34$ here (i.e. well below unity). The inset shows magnitude and phase of the first five PCA modes.}
	\label{fig3}
	\end{center}
\end{figure*}

We begin by considering a single realization with $M=4$ measurements and $N=64$ metamaterial elements per metasurface. The dipole moments and scene illuminations corresponding to the four learned metasurface configurations are displayed in Fig.~\ref{fig3}(a). The performance metric to evaluate the sensing ability is the average classification accuracy on a set of $10,000$ unseen samples. The confusion matrix in Fig.~\ref{fig3}(b) specifically shows how often a given digit is classified by the ANN as one of the ten options. The strong diagonal (corresponding to correctly identified digits) reflects the achieved average accuracy of $92.5\%$. Note that the diagonal entries are not expected to be equally strong (e.g. the $(1,1)$ entry is a bit stronger), since the test dataset does not include the exact same number of samples from each class. The off-diagonal entries of the confusion matrix are uniformly weak, so the ANN does not get particularly confused by any two classes.

\begin{figure*}
	\begin{center}
\includegraphics [width=\textwidth] {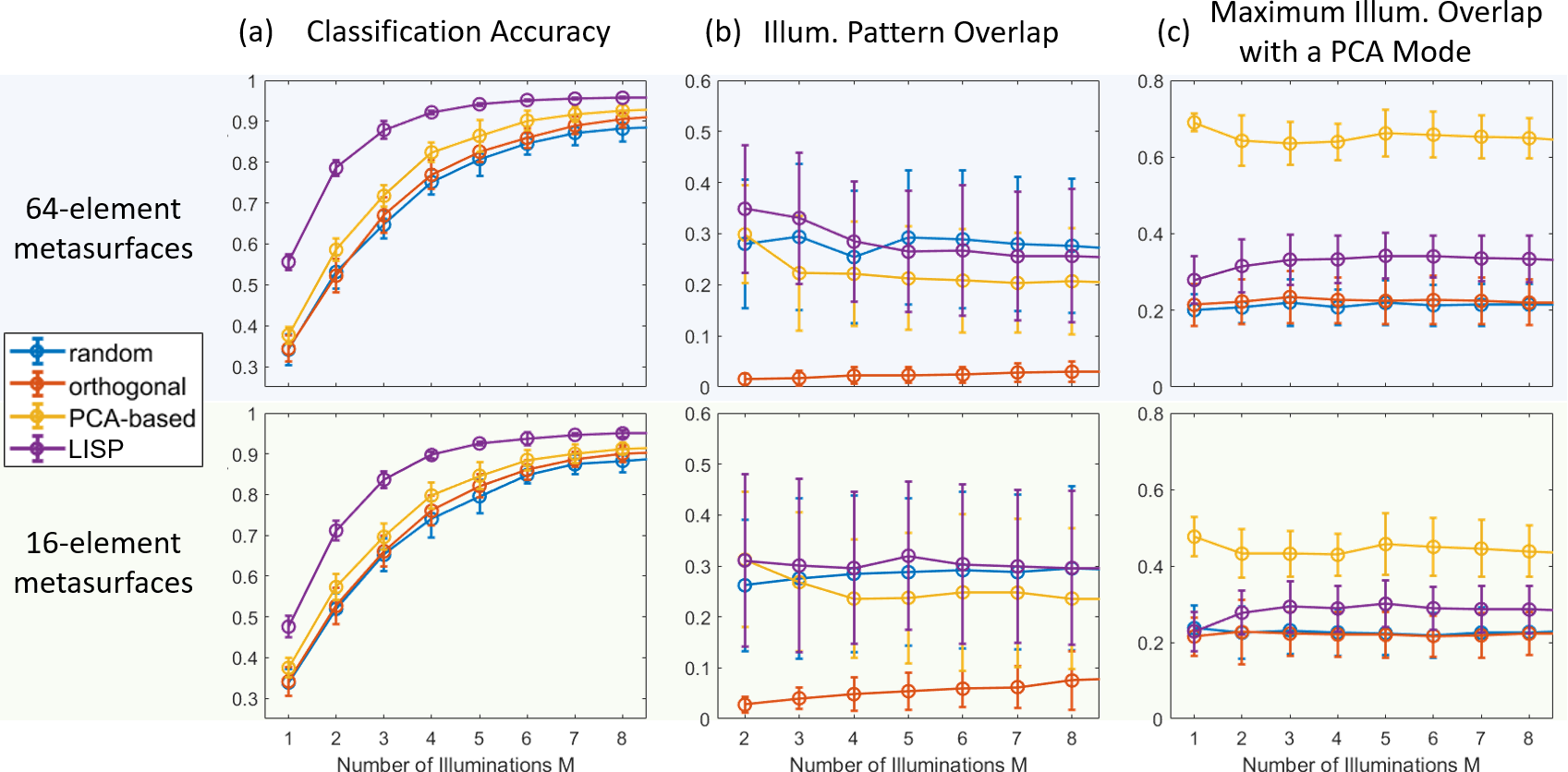}
	\caption{ (a) Comparison of the sensing performance with the 
	LISP illumination settings with three benchmarks: random, orthogonal and PCA-based scene illuminations. (b) Analysis of the mutual overlap between distinct scene illumination. (c) Analysis of the maximum overlap of the scene illuminations with a PCA mode.
	All data points are averaged over 40 realizations. Errorbars indicate the standard deviation.}
	\label{fig4}
	\end{center}
\end{figure*}

We now study the sensing performance more systematically for different values of $M$ and $N$. Since the ANN weights are initialized randomly before training, we conduct 40 realizations for each considered parameter combination. Averaging over realizations allows us to focus on the role of $M$ and $N$ without being sensitive to a given realization. Moreover, we can see the extent to which different realizations converge to similar results. 
In Fig.~\ref{fig4}(a) we contrast the sensing performance that we achieve by integrating a dynamic metasurface (consisting of $64$ or $16$ metamaterial elements) as physical layer into the ANN pipeline with the attainable performance if acquisition and processing are treated separately in schemes employing random, orthogonal or PCA-based scene illuminations. 

Random illuminations yield the worst performance out of the four considered paradigms. Orthogonal illuminations yield a marginal improvement over random illuminations as the number of measurements $M$ gets larger, since the non-overlapping illuminations can extract marginally more information about the scene. Yet the information that random and orthogonal illuminations encode in the measurements is not necessarily task-relevant. The PCA-based approach presents a noticeable performance improvement over generic illuminations but remains well below the attainable performance with learned optimal illumination settings obtained by integrating the analog layer into the ANN. Our LISP 
clearly outperforms the other three benchmarks. The PCA-based approach is quite sensitive to the number of tunable elements in the dynamic metasurface (keeping the same aperture size), since beam synthesis works better with more degrees of freedom. In contrast, using only $16$ instead of $64$ metamaterial elements yields almost identical sensing performance if the dynamic metasurface is integrated into the ANN pipeline. This suggests that using very few elements, and thus a very light hardware layer, our LISP 
can successfully perform sensing tasks. 

For $M\leq 5$, our scheme yields gains in accuracy of the order of $10\%$ which is a substantial improvement in the context of classification tasks \cite{AlexNet}; 
for instance, automotive security, where the number of scene illuminations is very limited, would be enhanced significantly if the recognition of, say, a pedestrian on the road could be improved by $10\%$. 
The performance using our learned illumination settings saturates around $M=5$ at $95\%$, meaning that we manage to extract all task-relevant information from a $28\times28$ scene with only $5$ measurements. 
The compression is enabled by the sparsity of task-relevant features in the scene. Yet our scene is not sparse in the canonical basis: our region of interest corresponds to the size of the metallic digits. Unlike traditional computational imaging schemes, the compression here comes from the discrimination between relevant and irrelevant information in a dense scene.
Our LISP 
thus achieves a significant dimensionality reduction by optimally multiplexing \textit{task-relevant} information from different parts of the scene in the analog layer. 
The dimensionality reduction brings about a double advantage with respect to timing constraints: taking fewer measurements takes less time, and moreover less data has to be processed by the digital layers. 
In our (not heavily optimized) ANN architecture, the computational burden of the first digital ANN layer is directly proportional to the number of measurements $M$. 
We thus believe that our scheme is very attractive in particular when real-time decisions based on wave-based sensing are necessary, notably in automotive security and touchless human-computer interaction. 
Moreover, a reduced processing burden can potentially avoid the need to outsource computations from the sensor edge to cloud servers via wireless links, mitigating associated latency and security issues \cite{hossain2015towards}.

A natural question that arises (albeit irrelevant for practical applications) is whether the other benchmark illumination schemes (random, orthogonal, PCA-based) will eventually, using more measurements, be able to perform as well as our LISP. 
For the $N=16$ case, we thus evaluated the average classification accuracy also for $M=N$ and $M=2N$. Only the LISP curve had saturated; the other benchmarks accurracies were still slightly improving between $M=N$ and $M=2N$ and were still somewhat below the LISP performance. Since the ``scene illumination'' $\mathbf{\mathcal{I}}_i(\mathbf{\zeta}) = \mathbf{E}^{\mathrm{TX}}_i(\mathbf{\zeta})  \cdot  \mathbf{E}^{\mathrm{RX}}_i(\mathbf{\zeta})$ depends on the configuration of two metasurfaces with $N$ tunable elements each, we expect that all schemes perform equally well only once $M\geq N^2$.

A striking difference in the performance fluctuations, evidenced by the error bars in Fig.~\ref{fig4}(a), is also visible. While the performance of our LISP 
does not present any appreciable fluctuations for $M \geq 4$, all other benchmark illumination schemes continue to fluctuate by several percent of classification accuracy. Our scheme's performance is thus reliable whereas any of the other benchmarks in any given realization may (taking the worst-case scenario) yield a classification accuracy several percent below its average performance. Performance reliability of a sensor is important for deployment in real-life decision making processes.

\subsection{Analysis of learned scene illuminations}

The inferior performance of orthogonal and PCA-based illuminations suggests that the task-specific learned LISP 
illuminations do not trivially correspond neither to a set of optimally diverse illuminations nor to the principal components of the scene. To substantiate this observation, we go on to analyze the scene illuminations in more detail. First, within a given series of $M$ illuminations, we compute the mutual overlap between different illuminations. In the following, we define the overlap $\mathcal{O}$ of two scene illuminations $A(\zeta)$ and $B(\zeta)$ as

\begin{equation}\label{overlap}
   \mathcal{O}(A,B) = \left| \frac{\int_{scene} A^\dag B \ \mathrm{d}\zeta}{\sqrt{\int_{scene} A^\dag A \ \mathrm{d}\zeta \ \int_{scene} B^\dag B \ \mathrm{d}\zeta}} \right| , %
\end{equation}

\noindent where $^\dag$ denotes the conjugate transpose operation. An example overlap matrix for $M=4$ is shown in Fig.~\ref{fig3}(c). We define the illumination pattern overlap as the mean of the off-diagonal elements (the diagonal entries are unity by definition).

In Fig.~\ref{fig4}(b) we present the average illumination pattern overlap for all four considered paradigms. For the case of orthogonal illuminations, the overlap is very close to zero, indicating that our inverse metasurface configuration design worked well. The inverse design works considerably better with $N=64$ as opposed to $N=16$, and of course the more illuminations we want to be mutually orthogonal the harder the task becomes. While radiating orthogonal wavefronts with the dynamic metasurface is not the best choice for sensing, it may well find applications in wireless communication \cite{del2019optimally}. 
For the case of random illuminations, the mutual overlap is constant at $28.5\%$ and independent of $N$. This is indeed the average overlap of two random complex vectors of length $10$, $10$ corresponding roughly to the number of independent speckle grains in the scene --- see Fig.~\ref{fig3}(a).

The mutual overlap of PCA-based illuminations, except for very low $M$, saturates around $21.5\%$ and $24.5\%$ for $N=64$ and $N=16$, respectively, and is hence lower than that of random illuminations. In principle, if beam synthesis worked perfectly, the PCA-based patterns should not overlap at all since PCA modes are orthogonal by definition. The more degrees of freedom $N$ we have, the better the beam synthesis works, and consequently the lower the mutual overlap of PCA-modes is. For the 
LISP 
scene illuminations, the average mutual overlap is comparable to that of random scene illuminations. We hence conclude that the diversity of the scene illuminations is not a key factor for the extraction of  \textit{task-relevant} information.

Next, we investigate to what extent the scene illuminations overlap with PCA modes. An example of the overlap with the first $25$ PCA modes is provided in Fig.~\ref{fig3}(d). In Fig.~\ref{fig4}(c), we present the average of the maximum overlap that a given illumination pattern has with any of the PCA modes. For random and orthogonal illuminations, irrespective of $N$ and $M$, this overlap is around $20\%$ and thus insignificant, as expected. For PCA-based illuminations we have performed beam synthesis to precisely maximize this overlap. We achieve $(65.0 \pm 1.9)\%$ with $N=64$ and $(44.1 \pm 1.4)\%$ with $N=16$. The ability to synthesize the PCA modes is thus very dependent on the number of metamaterial elements, and these results demonstrate that the PCA-based approach is suitable only for scenarios where $N$ is large. 
This observation is a further argument for the attractiveness of our approach in applications with very limited aperture size and tunability like automotive RADAR. In fact, since the PCA-based approach also requires an analytical forward model, one may as well choose the superior performance of our LISP proposal in any scenario where the PCA-based approach could be employed. Moreover, training with our approach is faster since all weights are optimized simultaneously, as opposed to first solving $M$ inverse design problems and then training the digital weights.

The overlap of the LISP illuminations with PCA modes is around $30\%$ and thus notably larger than for random or orthogonal illuminations but also notably lower than what can be achieved if one seeks PCA modes. Interestingly, the maximum overlap with a PCA mode is lower for $M=1$ and $M=2$. We conclude that the optimal illumination patterns identified by our ANN cannot simply be explained as corresponding to PCA modes, or to be a good approximation thereof. Notably for small $M$ this is not the case. During training, the ANN finds an optimal compromise taking the inner workings of the nonlinear digital layers as well as the physical layer constraints into account. The joint optimization of analog and digital layers provides substantially better performance than considering them separately and trying to anticipate useful illumination patterns.

We observe no significant difference in the performance across different metasurface realizations (i.e. different random locations of the metamaterial elements). The LISP scene illuminations overlap around $65\%$ across different realizations with the same metasurface, indicating that the optimization space contains numerous almost equivalent local minima. Remarkably, we never seem to get stuck in inferior local minima.

\subsection{Robustness}\label{sec_robustness}

Finally, we investigate how robust the sensing performance is outside the nominal conditions, i.e. over a given set of parameter variations. Here, we consider variations of the metamaterial elements' polarizability; due to fabrication tolerances of electronic components such as the PIN diodes the experimental polarizability is expected to differ across metamaterial elements by a few percent from the value extracted via full-wave simulation based on the element's design \cite{lauraprb}. With $M=5$ and $N=64$, we first train our LISP 
as before. Then, for each metamaterial element, we replace the true polarizability value $\alpha_{yy}$ by a different $\alpha'_{yy} = \alpha_{yy}\left( 1 + \epsilon \right)$, where $\epsilon$ is white noise; real and imaginary parts of $\epsilon$ are identically distributed with zero mean and standard deviation $\delta$, so the standard deviation of $\epsilon$ (the size of the cloud in the complex plane) is $\sqrt{2}\delta$. 

\begin{figure}[hhh]
	\begin{center}
\includegraphics [width=\columnwidth] {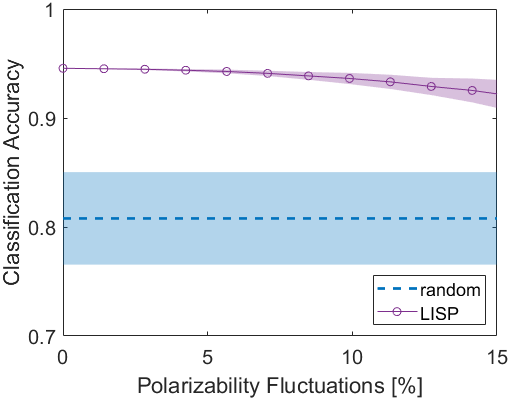}
	\caption{Robustness of sensing performance to fluctuations in the metamaterial elements' experimental polarizability. For a single realization with $N=64$ and $M=5$, the LISP 
	is trained with the expected polarizability value. Then, its performance is evaluated after adding different amounts of white noise to the metamaterial elements' polarizabilities. The purple curve and shaded area indicate average and standard deviation over $250$ realizations of the polarizability fluctuations. For reference, the blue curve and shaded area indicate the average performance with random metasurface configurations, see Fig.~\ref{fig4}(a), which is independent of polarizability fluctuations since it does not rely on optimized wavefronts.}
	\label{fig_robustness}
	\end{center}
\end{figure}

In Fig.~\ref{fig_robustness} we display the dependence of the sensing performance of our LISP 
scheme as a function of the polarizability fluctuations around the expected value. Mean and standard deviation over $250$ realizations of fluctuations are shown in purple.  Up to $5\%$ of fluctuations, the performance does not display any notable changes; around $10\%$ a very slight degradation of the performance is observed. The benchmark for comparison here is the case of using random illuminations since these do not rely on carefully shaped wavefronts and are thus not affected by experimental polarizability values different from the expected ones. Compared to the performance with random illuminations, our scheme is still clearly superior even for unrealistically high fluctuations on the order of $15\%$. These results suggest that realistic deviations from the expected polarizability values of the metamaterial elements, due to fabrication tolerances or other neglected effects, are by no means detrimental for our scheme.

\section{Outlook}

In spite of the encouraging robustness results shown in the previous section~\ref{sec_robustness}, it is worth considering how one would deal with significant fabrication inaccuracies in experimental metasurfaces. 
Should one or several parameters of the metasurface design, such as polarizabilities or locations of the metamaterial elements, turn out to significantly vary due to fabrication issues, an additional calibration step could easily learn the actual parameter values of a given fabricated metasurface. The spirit of this procedure is similar to the way in which we achieved beam synthesis (see Section~B.3 of the Supplemental Material) 
using only the analog part of our pipeline. The parameters to be determined are declared as weight variables to be learned during training and initialized with their expected values. Then experimentally the radiated fields for a few different metasurface configurations are measured and a cost function is defined to minimize during training the difference between the measured and expected radiated fields. By the end of the optimization, a set of parameter values will have been learned that optimally matches the experiment.

Furthermore, for practical applications it may be of interest to achieve a certain robustness to fluctuations in the calibration parameters such as geometrical details of the experimental setup \cite{mutapcic2009robust,sniedovich2016statistical,satat2017object}. By including random realizations of major calibration parameters within the expected range of fluctuation during the training, the network can learn to be invariant to these variations \cite{satat2017object}. Ultimately, this enables the transfer of knowledge learned on synthesized data to real-life setups without additional training measurements. For a specific task such as hand gesture recognition, synthetic scenes can easily be generated with appropriate 3D modelling tools.

In future implementations, it may be worthwhile to include additional measurement and processing constraints in the LISP, such as phaseless measurements and few-bit processing \cite{sze2017efficient}, to lower the hardware requirements further. With more advanced forward models beyond the first Born approximation, quantitative imaging may become possible, for instance of the dielectric constant in contexts ranging from the detection of breast cancer to the inspection of wood \cite{fear2002confocal,boero2018microwave}.
In practice, incremental learning techniques \cite{yoon2017lifelong,castro2018end} may enable one to adapt the LISP efficiently to new circumstances without retraining from scratch, for instance if an additional class is to be recognized.

Finally, a conceptually exciting question for future research is how we can conceive a LISP 
that is capable of taking real-time on-the-fly decisions about the next optimal scene illumination, taking into account the available knowledge from previous measurements \cite{realtime}.

\section{Conclusion}

In summary, we have shown that integrating the physical layer of a wave-based sensor into its artificial-neural-network (ANN) pipeline substantially enhances the ability to efficiently obtain task-relevant information about the scene. The jointly optimized analog and digital layers optimally encode relevant information in the measurements, converting data acquisition simultaneously into a first layer of data processing. A thorough analysis of the learned optimal illumination patterns revealed that they cannot be anticipated from outside the ANN pipeline, for instance by considerations to maximize the mutual information or based on principal scene components, highlighting the importance of optimizing a unique analog-digital pipeline. 

As concrete example, we considered the use of dynamic metasurfaces for classification tasks which are highly relevant to emerging concepts of ``context-awareness'', ranging from real-time decision making in automotive security via gesture recognition for touchless human-device interactions to fall detection for smart health care of elderly. The dynamic metasurfaces, thanks to their inherent analog multiplexing and structural compactness, are poised to play an important role in these applications. Moreover, the sub-wavelength nature of the embedded metamaterial elements enabled us to formulate a very compact analytical forward model based on a coupled-dipole description that we then combined with the machine learning framework of our Learned Integrated Sensing Pipeline. 
Several traditional inverse-design hurdles like binary constraints on the analog weights and coupling between metamaterial elements were cleared with ease in our scheme. 
In addition to a substantially higher classification accuracy, we observed that our scheme is more reliable (very low performance fluctuations across realizations) and very robust (against fabrication inaccuracies).

The ability to very efficiently extract task-relevant information greatly reduces the number of necessary measurements as well as the amount of the data to be processed by the digital layer, which is very valuable in the presence of strict timing constraints as found in most applications. We expect our work to also trigger interesting developments in other areas of wave physics, notably optical and acoustic wavefront shaping \cite{rotter2017light,ma2018shaping}. 
\bigskip

\section*{Acknowledgments}

M.F.I., A.V.D. and D.R.S. are supported by the Air Force Office of Scientific Research under award number FA9550-18-1-0187.

\bigskip

The project was initiated and conceptualized by P.d.H. with input from M.F.I. and R.H. The project was conducted and written up by P.d.H. All authors contributed with thorough discussions.

\providecommand{\noopsort}[1]{}\providecommand{\singleletter}[1]{#1}%

\clearpage

\section*{SUPPLEMENTAL MATERIAL}
\renewcommand{\thefigure}{S\arabic{figure}}
\renewcommand{\theequation}{S\arabic{equation}}
\setcounter{equation}{0}
\setcounter{figure}{0}

For the interested reader, here we provide numerous additional details that complement the
manuscript and provide further illustrations. This document is organized as follows:

A. Detailed Description of the Physical Layer.

\hspace{10pt} 1. Dipole Moments for Array of Interacting Dipoles.

\hspace{10pt} 2. Visualization of Dipole Interaction.

\hspace{10pt} 3. Tuning Mechanism.

\hspace{10pt} 4. Propagation to Scene.

B. Details of Artificial Neural Network Algorithm and Parameters.

\hspace{10pt} 1. Training Algorithm and Parameters.

\hspace{10pt} 2. Imposing a Discrete Distribution of Weights.

\hspace{10pt} 3. Constrained Beam Synthesis / Wavefront Shaping.

C. Taxonomy of Illumination Strategies in Wave-Based Sensing.

\subsection{Detailed Description of the Physical Layer}

In this section we provide, for completeness, the equations used (i) to compute the dipole moments and (ii) to describe the propagation to the scene. We refer the interested reader to Ref.~\cite{laura2018} and Ref.~\cite{guy2015}, respectively, where these expression are derived in a more general form. 

\subsubsection{Dipole Moments for Array of Interacting Dipoles}\label{dmaip}

The goal of this section is to compute the dipole moments of each radiating element for a given configuration to then use them to compute the radiated field in the scene as detailed in Section~\ref{propagscene}.

The metamaterial elements can be described as magnetic dipoles due to their small size compared to the wavelength. Using surface equivalent principles \cite{lauraprb}, an equivalent polarizability tensor $\rttensor{\alpha}$ for a given metamaterial element can be extracted. The metamaterial's dipole moment $\mathbf{m}$ is then determined as

\begin{equation}\label{eq1}
  \mathbf{m} (\mathbf{r}) = \rttensor{\alpha} \ \mathbf{H^{loc}} (\mathbf{r}) ,%
\end{equation}\label{dipole_polariz_field}

\noindent where $\mathbf{H_{loc}}$ is the local field at the metamaterial's position $\mathbf{r}$.

For the example metamaterial element shown in the inset of Fig.~1(a) in the main text (a tunable cELC-resonator \cite{CELC,TimCELC}), at 
the working frequency of $f_0 = 10\ \mathrm{GHz}$ (wavelength $\lambda_0 = 0.03 \ \mathrm{m}$), only the $\alpha_{yy}$ component of the polarizability tensor is significant \cite{laura2018}. This simplifies Eq.~\ref{eq1} to

\begin{equation} \label{eq2}
  m_y (\mathbf{r}) = \alpha_{yy} \ H^{loc}_y (\mathbf{r}) %
\end{equation}\label{dipole_polariz_field_scal}

\noindent for the setup we consider. 

The crux now lies in computing the local field $H^{loc}_y (\mathbf{r}) $ exciting a given metamaterial element. $H^{loc}_y (\mathbf{r}) $ is a superposition of the cylindrical wave feeding the waveguide, $H^{feed}_y (\mathbf{r})$, and the fields $H^{interact}_y (\mathbf{r}) $ radiated from the other metamaterial elements:

\begin{equation}
  H^{loc}_y (\mathbf{r})  = H^{feed}_y (\mathbf{r}) + H^{interact}_y (\mathbf{r}) . %
\end{equation}

The analytical expression for $H^{feed}_y (\mathbf{r})$ reads

\begin{equation}\label{feedwave}
  H^{feed}_y (\mathbf{r}) = \frac{\mathrm{i} I_e k}{4}  \ \mathrm{H^{(2)}_1}\! \left( k | \mathbf{r} - \mathbf{r_0} | \right) \ \mathrm{sin}(\theta), %
\end{equation}

\noindent where $\mathrm{i}$ is the imaginary unit, $I_e$ is the amplitude of the electric line source generating the feed wave (taken to be $1 \ \mathrm{A}$ in the following), $k = 2\pi f_0 / c$ is the propagation constant of the fundamental mode inside the waveguide, $\mathrm{H^{(2)}_1}$ is a first order Hankel function of the second kind, $\mathbf{r_0}$ is the position of the source, and $\theta$ is the circumferential angle around the source measured from the $y$-axis. The space inside the waveguide we consider is empty (not filled with a substrate).

The field $H^{interact}_y (\mathbf{r_i}) $ exciting the $i$th dipole due to the presence of other dipoles can be related to the Green's function $G_{yy}(\mathbf{r_i},\mathbf{r_j})$ between the element under consideration at position $\mathbf{r_i}$ and the $j$th element at position $\mathbf{r_j}$ as

\begin{equation}\label{Hinteract}
  H^{interact}_y (\mathbf{r_i}) = \sum_{i\neq j} {  G_{yy}(\mathbf{r_i},\mathbf{r_j}) \ m_y(\mathbf{r_j}) }. %
\end{equation}

The self-interaction term, i.e. $i=j$, is included in the definition of effective polarizability, as detailed in Section~II of Ref.~\cite{lauraprb}. Note that Eq.~\ref{eq2} and Eq.~\ref{Hinteract} are coupled via the $m_y(\mathbf{r_j})$ term. To compute the dipole moments $m_y(\mathbf{r_j})$, we thus have to solve a system of coupled equations. First, rearranging both to yield an expression for $H^{loc}_y (\mathbf{r_i})$ yields

\begin{subequations}\label{HlocLHS}
\label{eq:whole}
\begin{eqnarray}
H^{loc}_y (\mathbf{r_i}) =  \alpha_{yy}^{-1} \ m_y(\mathbf{r_i}) ,\label{subeq:2}
\end{eqnarray}
\begin{equation}
H^{loc}_y (\mathbf{r_i}) = H^{feed}_y (\mathbf{r_i}) + \sum_{i\neq j} {  G_{yy}(\mathbf{r_i},\mathbf{r_j}) \ m_y(\mathbf{r_j}) } .\label{subeq:1}
\end{equation}
\end{subequations}

\noindent Next, we rewrite Eq.~\ref{HlocLHS} in matrix form:

\begin{equation}\label{matrix_form}
  \mathbf{A}\mathbf{M} = \mathbf{F} + \mathbf{G} \mathbf{M}  , %
\end{equation}

\noindent where $\mathbf{A}$ is a diagonal matrix whose diagonal entries are $\alpha_{yy}^{-1}$ (all metamaterial elements are identical), $\mathbf{M}$ is a vector containing the sought-after dipole moments, $\mathbf{G}$ is a matrix whose entry $(i,j)$ corresponds to the Green's function between elements at positions $\mathbf{r_i}$ and $\mathbf{r_j}$ (the diagonal entries are set to zero since the self-interaction terms are incorporated into the effective polarizabilities in $\mathbf{A}$), and $\mathbf{F}$ is a vector containing the fields due to the feeding wave at the element positions.

Solving Eq.~\ref{matrix_form} for $\mathbf{M}$ then yields

\begin{equation}\label{solution_for_M}
  \mathbf{M} = \{ \mathbf{A} - \mathbf{G} \}^{-1} \mathbf{F}   . %
\end{equation}

Now, the remaining step is to identify an expression for the inter-element Green's function $G_{yy}(\mathbf{r_i},\mathbf{r_j})$. The inter-element coupling consists of two components --- interactions via the waveguide (WG) and interactions via free space (FS):

\begin{equation}
  G_{yy}(\mathbf{r_i},\mathbf{r_j}) = G^{WG}_{yy}(\mathbf{r_i},\mathbf{r_j}) + G^{FS}_{yy}(\mathbf{r_i},\mathbf{r_j}). %
\end{equation}

\noindent The waveguide interaction component reads

\begin{equation}\label{waveguideinteraction}
   G^{WG}_{yy}(\mathbf{r_i},\mathbf{r_j}) = - \frac{\mathrm{i} k^2}{8h}\left( \mathrm{H^{(2)}_0}\! \left( k R \right) - \mathrm{cos}(2\phi)\  \mathrm{H^{(2)}_2}\! \left( k R \right) \right), %
\end{equation}

\noindent where $R=| \mathbf{r_i} - \mathbf{r_j} |$ and $\phi$ is the angle of the vector from $ \mathbf{r_i}$ to $\mathbf{r_j}$. The free space interaction \footnote{The expression for $G^{FS}_{yy}$ can be obtained from its definition  
\begin{equation*}\label{xx}
   G^{FS}_{yy}(\mathbf{r_i},\mathbf{r_j}) =  \left(k^2 + \frac{\partial ^2}{\partial y_i^2} \right) \frac{\mathrm{e}^{-\mathrm{i}k|\mathbf{r_i}-\mathbf{r_j}|}}{4\pi |\mathbf{r_i}-\mathbf{r_j}|}  %
\end{equation*}
\noindent or via the Dyadic Green's Function \cite{jackson,Langmuir}
\begin{equation*}\label{dya}
    \rttensor{G} (\mathbf{r_i},\mathbf{r_j}) = g(R) \left( \left( \frac{3}{R^2} - \frac{3\mathrm{i}k}{R} - k^2 \right)\hat{r}\hat{r} + \left( k^2+ \frac{\mathrm{i}k}{R}-\frac{1}{ R^2} \right)\rttensor{I} \right),
\end{equation*}

\noindent where $\hat{r}$ is a unit vector parallel to $\mathbf{r_i}-\mathbf{r_j}$ and $\hat{r}\hat{r}$ denotes an outer product. 
}
is given by

\begin{equation}\label{Gfs}
   G^{FS}_{yy}(\mathbf{r_i},\mathbf{r_j}) =  g(R) \left(  k^2 + \frac{\mathrm{i}k}{R} - \frac{1+k^2 \Delta_y^2 }{R^2} - \frac{3\mathrm{i} k\Delta_y^2}{R^3} + \frac{3\Delta_y^2}{R^4}  \right) , %
\end{equation}

\noindent where $\Delta_y = y_i -y_j$ and $g(R) = 2 \frac{\mathrm{e}^{-\mathrm{i} kR} }{4\pi R}$. The factor $2$ in the expression for $g(R)$ originates from self-images of the elements due to the waveguide's metallic upper plate.

\subsubsection{Visualization of Dipole Interaction}

In this section we provide a visualization of the importance of accounting for the coupling between different metamaterial elements. To that end, we first compute for a random configuration of the metasurface shown in Fig.~1(a) of the main text  
the corresponding dipole moments for each metamaterial element. We compare this with a linear superposition of the individual elements, i.e. omitting the presence of other dipoles in calculating each element's dipole moment.

\begin{figure}[h]
	\begin{center}
\includegraphics [width=\columnwidth] {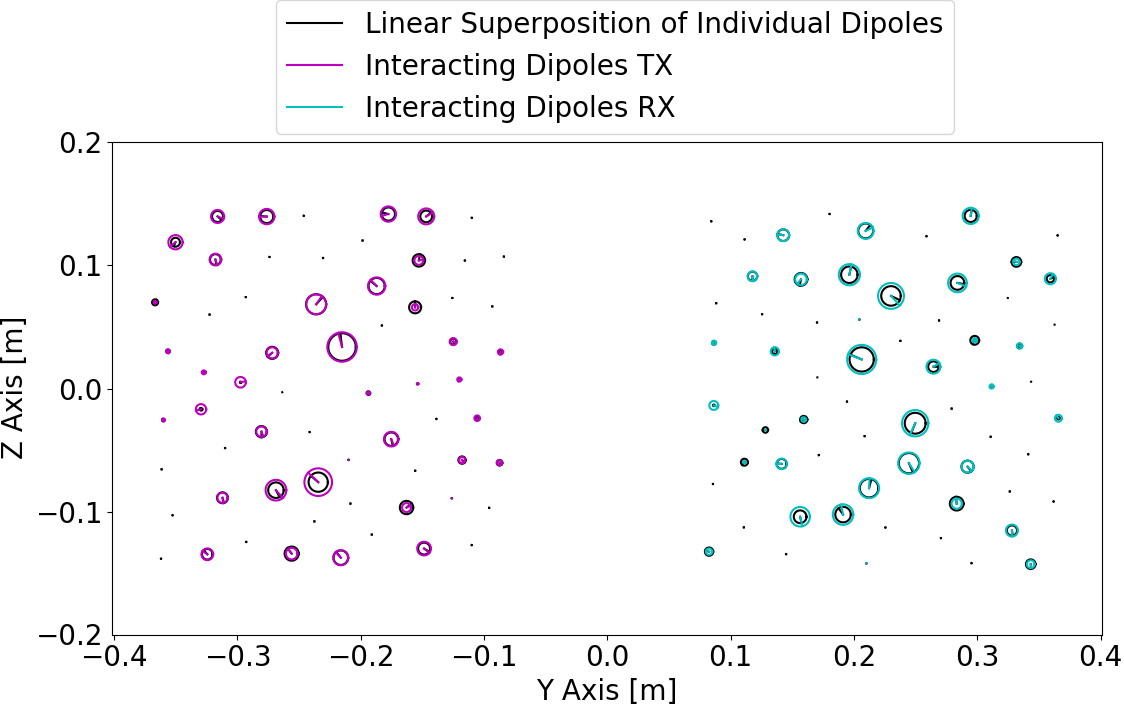}
	\caption{Comparison of the dipole moments $m_y$ computed for each metamaterial element, without (black) and with (purple/cyan) accounting for the coupling between elements. The dipole moments are represented as phasors, with the radius of the circle being proportional to their amplitude, and the line segment showing their phase. The circles are centered on the physical location of each metamaterial element.  } 
	\label{VisCoupl}
	\end{center}
\end{figure}

The comparison in Fig.~\ref{VisCoupl} clearly shows the non-negligible difference between the two results, highlighting the importance of accounting for the coupling between metamaterial elements.

\subsubsection{Tuning Mechanism}\label{Tuning Mechanism}

So far, the above calculations are for a device that is not reconfigurable, i.e. all included dipoles are radiating energy. For our scheme based on reconfigurability it is hence crucial to add a description of the tuning mechanism. Any phase or amplitude tuning is relative to the element's polarizability $\alpha_{yy}$; for instance, the tuning states ``on'' and ``off'' that are accessible with a metamaterial element tuned via PIN-diodes connected across the cELC's capacitive gaps \cite{TimCELC} correspond to multiplying the polarizability by $1$ or $0$. If the polarizability is zero, the element is not radiating any energy and is hence essentially non-existant. This simple picture neglects ohmic losses associated with the PIN diodes when they are in conducting mode (i.e. the metamaterial element is not radiating); these losses are usually small but could be accounted for in any given practical implementation.

Consider a general tuning state described via a vector $\mathbf{T}$ whose $j$th entry $t_j$ corresponds to the tuning of the $j$th element. The element's polarizability is then $t_j\alpha_{yy}$. Correspondingly, the $j$th diagonal entry of $\mathbf{A}$ becomes $\{ t_j\alpha_{yy} \}^{-1}$. In practice, to avoid issues related to division by zero, we use $\{ t_j\alpha_{yy} + \delta \}^{-1}$ with $\delta$ being four orders of magnitude smaller than $\alpha_{yy}$.

\begin{figure}[h]
	\begin{center}
\includegraphics [width=\columnwidth] {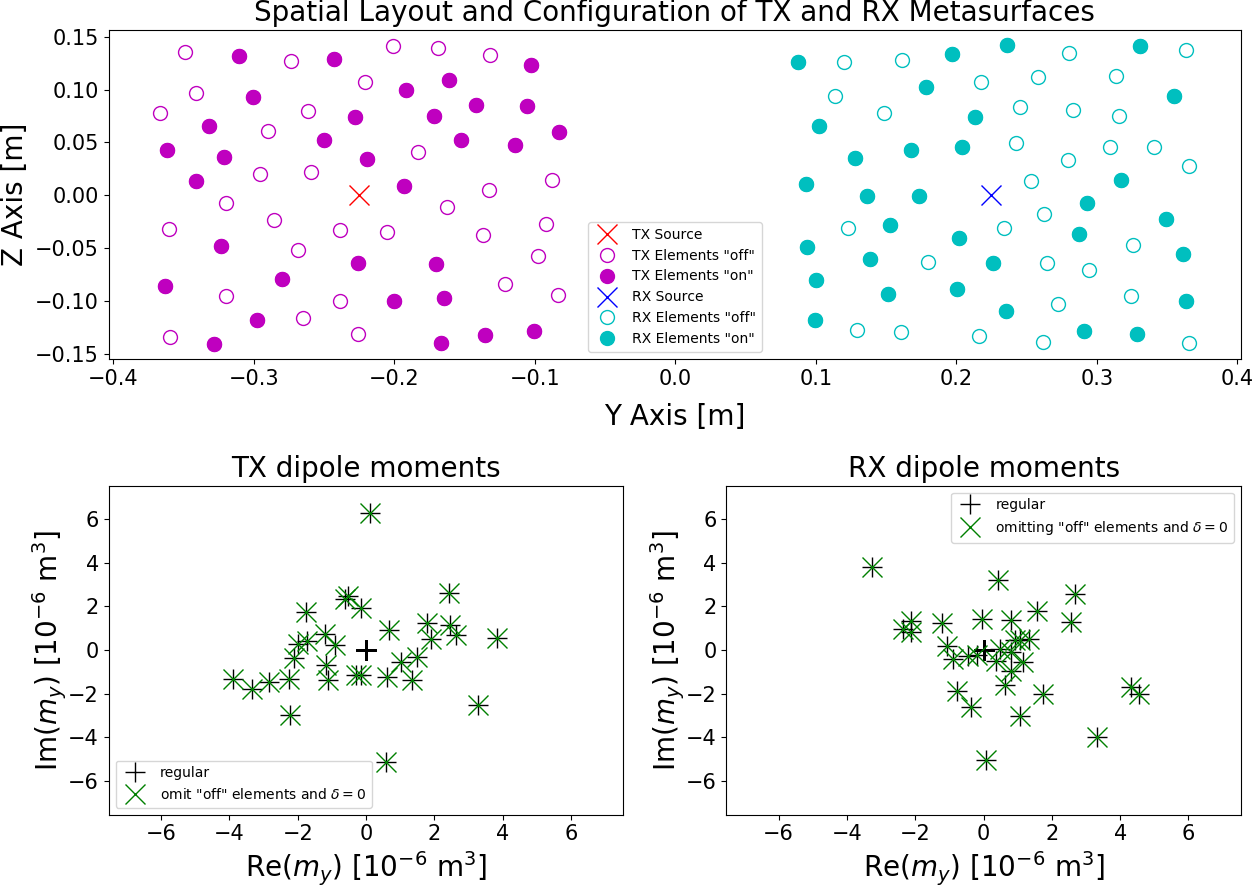}
	\caption{Sanity check of the tuning mechanism considering metasurfaces with 64 metamaterial elements. The top figure shows the spatial layout of the considered metasurfaces, as well as the chosen (random) configurations: filled elements are ``on'', unfilled elements are ``off''. The bottom row contrasts the computed dipole moments for the two metasurfaces, once using $\{ t_j\alpha_{yy} + \delta \}^{-1}$ as entries of the diagonal of $\mathbf{A}$ (black) and once omitting the ``off'' elements and using $\{\alpha_{yy}\}^{-1}$ as entries of the diagonal of $\mathbf{A}$ (green). Excellent agreement is seen.  } 
	\label{SsCT}
	\end{center}
\end{figure}

A simple sanity check to confirm the above procedure consists in computing the magnetic dipole moments corresponding to a binary (random) configuration $\mathbf{T}$ and comparing the result to that obtained if elements in the ``off'' state are simply omitted. Ideally the same magnetic dipole moments are computed for the ``on'' elements in both cases. This is indeed the case, as shown in Fig.~\ref{SsCT}.

\subsubsection{Propagation to Scene}\label{propagscene}

Having found a description of the transceiver ports in terms of dipole clusters, we can now proceed with computing the fields illuminating the scene for a given choice of physical layer weights. The considered setup is shown in Fig.~1(b) in the main text. 
To that end, we compute the superposition of the radiated fields from all dipoles in the scene as 

\begin{equation}\label{e_tx}
   \mathbf{E^{TX}}(\mathbf{\zeta}) = \frac{-\mathrm{i}\omega \mu_0}{4\pi} \sum_j \left( (\mathbf{m}(\mathbf{r_j}) \times \hat{\mathbf{\rho_j}} ) \left( \frac{-\mathrm{i}k}{\Gamma_j} - \frac{1}{\Gamma_j^2} \right) \mathrm{e}^{-\mathrm{i}k\Gamma_j} \right), %
\end{equation}

\noindent where $\zeta$ is a position in the scene, $\omega=2\pi f_0$, $\Gamma_j = |\mathbf{\zeta} - \mathbf{r_j}|$, $\hat{\mathbf{\rho_j}}$ is a unit vector parallel to $\mathbf{\zeta}-\mathbf{r_j}$ and $\mathbf{m}(\mathbf{r_j})$ is the dipole moment of the $j$th transmit (TX) dipole located at position $\mathbf{r_j}$.

Given that only the $y$-component of $\mathbf{m}(\mathbf{r_j})$ is significant in our case, we can evaluate $\mathbf{m}(\mathbf{r_j}) \times \hat{\mathbf{\rho_j}}$ as follows:

\begin{equation}\label{e_tx}
   \begin{bmatrix}
           0 \\
           m_{y}(\mathbf{r_j}) \\
           0
         \end{bmatrix} \times  \begin{bmatrix}
           (x_\zeta - x_{\mathbf{r_j}}) / \Gamma_j \\
           (y_\zeta - y_{\mathbf{r_j}}) / \Gamma_j \\
           (z_\zeta - z_{\mathbf{r_j}}) / \Gamma_j 
         \end{bmatrix}  
         = \frac{m_{y}(\mathbf{r_j})}{\Gamma_j} \begin{bmatrix}
           z_\zeta - z_{\mathbf{r_j}}  \\
           0 \\
           x_{\mathbf{r_j}}- x_\zeta  
         \end{bmatrix},
\end{equation}

\noindent where $x_\zeta$ denotes the $x$-component of $\mathbf{\zeta}$ etc. Hence $\mathbf{E^{TX}}$ has non-zero components along $z$ and $x$, with $z$ being the dominant one.

These fields are then reflected by the scene and the receive (RX) port captures the reflected fields. The entering fields at the RX port are simply the time-reversed version of an expression analogous to that given for the exiting fields at the TX port in Eq.~\ref{e_tx}.

Note that the physical layer weights are implicitly accounted for in Eq.~\ref{e_tx} via the dipole moments as discussed in Section~\ref{Tuning Mechanism}.

We can thus compute the $i$th measurement (a complex-valued scalar), obtained with the $i$th pair of TX/RX physical weights, as

\begin{equation}\label{measurement}
   g_i \propto \int_{scene}  \mathbf{E}^{\mathrm{TX}}_i(\mathbf{\zeta})  \cdot  \mathbf{E}^{\mathrm{RX}}_i(\mathbf{\zeta}) \ \sigma(\mathbf{\zeta}) \ \mathrm{d}\mathbf{\zeta}, %
\end{equation}

\noindent where $\mathbf{\sigma}$ is the scene reflectivity. Here, we choose to work with a reflectivity corresponding to the gray-scale pixel values of an image from the MNIST dataset of handwritten digits \cite{mnist}. In general, Eq.~\ref{measurement} relies on a first Born approximation which is appropriate for our flat scene (no variation in the $x$-direction).

\subsection{Details of Artificial Neural Network \\Algorithm and Parameters}

In this section, we detail the implementation of our method using a low-level API of the open source machine learning library TensorFlow \cite{tensorflow}. The ANN architecture has been described in the main text. We use the float32 and complex64 data types for real and complex valued variables, respectively.

Since the metamaterial elements are separated by at least a wavelength, the arguments of the Hankel functions in Eq.~\ref{feedwave} and Eq.~\ref{waveguideinteraction} are always above unity ($kR \geq 2\pi$). We thus evaluate these Hankel functions with 
the following large-argument approximation:

\begin{equation}\label{measurement}
   \mathrm{H}^{(2)}_\eta (\xi) \xrightarrow[\xi > 1]{} \sqrt{\frac{2}{\pi\xi}} \ \mathrm{e}^{-\mathrm{i} \left( \xi - \frac{\eta \pi}{2} - \frac{\pi}{4} \right)} .%
\end{equation}

We initialize the bias variables ($b_p^{(1)}$ and $b_c^{(2)}$ in Fig.~2 of the main text) as zero. Weight variables are initialized with a truncated normal distribution \footnote{The generated values follow a normal distribution with specified mean and standard deviation, except that values whose magnitude is more than 2 standard deviations from the mean are dropped and re-picked.}; mean and standard deviation of the latter are zero and $0.12$, respectively, for the digital weight variables ($w_{p,i}^{(1)}$ and $w_{c,p}^{(2)}$ in Fig.~2 of the main text) and zero and $0.2$, respectively, for the physical weight variables $t_{j,i}$.

\subsubsection{Training Algorithm and Parameters}\label{training}

We use the MNIST dataset of handwritten digits \cite{mnist} which consists of $60000$ samples of $28\times28$ gray-scale images of handwritten digits from $0$ to $9$ and $10000$ further such samples to test the learned network. For training, we split the training dataset and use $15\ \%$ for validation purposes. Using a batch size of $n_{batch} = 100$ and the Adam method for stochastic optimization \cite{adam} with a step size of $10^{-3}$, we train the weight variables of the network based on the training dataset, using the cross entropy between the known and predicted labels as error metric to be minimized. Every $50$ epochs we compute the accuracy achieved on the validation dataset. We define a patience parameter to avoid overlearning on the training dataset: if the validation accuracy has not improved for seven consecutive times, we stop training. Finally, we evaluate the achieved accuracy on the completely unseen test dataset. This is the accuracy we report in the main text.

Since our measurements are complex-valued, we stack real and imaginary parts of the $M$ measurements in a $1\times2M$ real-valued vector $\mathcal{M}$; then we feed $\mathcal{M}$ into the first fully connected layer. A subtle but crucial detail is the need to normalize $\mathcal{M}$. If its standard deviation is several orders of magnitude above or below unity, training the weights of the fully connected layers may take much longer or be unfeasible, in particular in combination with the temperature parameter (see below). Hence, we first identify a normalization constant as the average of the standard deviation of $\mathcal{M}$ obtained for ten different random TX and RX masks (obeying the chosen constraint such as amplitude-binary) using the training dataset. We then always divide $\mathcal{M}$ by that fixed normalization constant in the subsequent training, validation and testing. In other words, the normalization constant is fixed once and for all based on the training dataset. 

The employed activation functions are defined for real-valued inputs as follows. A ``rectified linear unit'' (ReLu) activation function returns $0$ if its argument $x$ is negative:

\begin{equation}\label{ReLu}
   f_{ReLu}(x) = \mathrm{max}(x,0).
\end{equation}

\noindent A SoftMax activation normalizes a $p$-element vector $\mathbf{Y}=[Y_1,Y_2,\dots,Y_p]$ into a probability distribution consisting of $p$ probabilities whose sum yields unity:

\begin{equation}\label{SoftMax}
   f_{SoftMax}(Y_p) = \frac{\mathrm{e}^{Y_p}}{\sum_p {\mathrm{e}^{Y_p}}}.
\end{equation}

\subsubsection{Imposing a Discrete Distribution of Weights}

In this section, we provide the technical details of how we restrict the physical layer weights to be chosen from a list $\mathcal{S}$ of $s$ predefined discrete values. For instance, the reconfiguration mechanism of the metamaterial elements may leverage PIN-diodes such that a given element is either resonant or not at the working frequency $f_0$, which corresponds to a binary on/off amplitude modulation: $\mathcal{S}=\{0,1\}$. Our procedure can also be directly applied to gray-scale tuning, for instance in the case of a phased-array with $2$-bit phase modulation we would use $\mathcal{S}=\{1,\mathrm{i},-1,-\mathrm{i}\}$. (Preliminary tests showed little gains in the sensing performance with gray-scale tuning compared to amplitude-binary tuning.)

The ANN weight variables are, however, trained via back-propagating errors \cite{backpropNature} which relies on computing gradients. This approach is thus, at first sight, incompatible with variables that can only take a few discrete values. An elegant solution to this apparent problem consists in using a temperature parameter \cite{chakrabarti}. The general idea is to start with a continuously-distributed weight variable and to gradually force its distribution to become more and more discrete over the course of the iterations. By the end of the training, the weights are then effectively discrete and chosen from $\mathcal{S}$.

We define a scale factor $\beta$ that increases according to a quadratic schedule with the number of iterations $T$:

\begin{equation}
   \beta (T) = 1+(\gamma T)^2, %
\end{equation}

\noindent where we use $\gamma = 0.0005$. For every variable $t_i$ to be trained, we introduce an $s$-element vector $\mathbf{V_i} = [t_{i,1} , t_{i,2}, \dots, t_{i,s}]$. We then multiply the absolute value of $\mathbf{V_i}$ with the scale factor $\beta$ and apply a SoftMax operation (see Eq.~\ref{SoftMax}): $ \mathbf{V'_i} = f_{SoftMax}(\beta | \mathbf{V_i}|)$. We go on to define $t_i$ as a sum of its possible values $ S_j\in\mathcal{S}$ weighted by $t'_{i,j}$ (the entries of $ \mathbf{V'_i}$): 
$t_i= \sum_j t'_{i,j} S_j$. 
The entries of $\mathbf{V_i}$ are declared as weight variables and are trained via error back-propagation. The SoftMax function in combination with the gradually increasing scale factor ensures that eventually $ \mathbf{V'_i}$ ends up having one unity entry and the remaining entries become zero. We thus gradually transition from a continuous to a discrete distribution. To ensure that the $t_i$ variables are indeed binary after training, we defer checking the patience parameter until $\mathrm{min}(t_{i,j},1-t_{i,j})_{i,j}$ is below $0.0001$.

\subsubsection{Constrained Beam Synthesis / Wavefront Shaping}\label{sec_beamsynthesis}

This section is 
motivated by the wish to compare the performance of our Learned Integrated Sensing Pipeline (LISP) not only with random illuminations but also with orthogonal and PCA-based illuminations. 
For PCA-based illuminations, $M$ inverse design problems of beam synthesis have to be solved to identify metasurface configurations whose scene illuminations approximate as closely as possible the first $M$ PCA modes. For orthogonal scene illuminations, a set of $M$ configurations has to be identified such that their scene illuminations' overlap as little as possible. Note that we do not impose a specific orthogonal basis (such as the Hadamard basis); the obtained illuminations with minimal mutual overlap will thus still look speckle-like. Both of these inverse design problems are notoriously difficult due to (i) the binary constraint and (ii) the inter-element coupling \cite{boyd2004convex}. Traditional approaches to tackle such NP-hard combinatorial optimization problems (even without the coupling constraint) include Gerchberg-Saxton (GS) algorithms \cite{fienup1982phase} and semidefinite programming tools.

Here, we choose a much simpler approach resembling ``adjoint''-based methods \cite{jensen2011topology,chung2019high}.
We take the analog part of our pipeline in Fig.~2 of the main text and define a new cost-function based upon $\mathcal{I}_{z,i} =  {E}^{\mathrm{TX}}_{z,i}(\mathbf{\zeta})  \cdot  {E}^{\mathrm{RX}}_{z,i}(\mathbf{\zeta})$ (since $z$ is the dominant component). More specifically, we either maximize for a single mask the overlap with a principal scene component or minimize for a series of masks the average mutual overlap of the scene illuminations.

In order to identify $M$ illuminations with minimal mutual overlap, we first compute the $M\times M$ overlap matrix $\Omega$ whose $(i,j)$th entry is $\Omega_{i,j} = \mathcal{O}(\mathcal{I}_{z,i}(\zeta),\mathcal{I}_{z,j}(\zeta))$, where $\mathcal{O}$ is the overlap function as defined in Eq.~6 of the main text. Next, we compute the average mutual overlap as the average of the off-diagonal entries of $\Omega$ and use that as cost function to be minimized: $\langle \Omega_{i,j}\rangle_{i\neq j}$.

In order to identify an illumination pattern that matches a given PCA mode $\mathcal{P}$ as closely as possible, we compute the overlap of the scene illumination $\mathcal{I}_z$ with $\mathcal{P}$ and use its inverse as cost function to be minimized: $1/\mathcal{O}(\mathcal{I}_z,\mathcal{P})$. Minimizing this cost function will maximize the resemblance of $\mathcal{I}_z$ to $\mathcal{P}$.

Using the temperature parameter as before ensures that during training the weights are carefully driven towards a binary distribution. This simple approach to deal with constraints and even coupling effects, only requiring the formulation of an analytical forward model, may also prove useful in other \textit{constrained} physical inverse design problems, from nano-photonic inverse design \cite{piggott2015inverse} and optical wavefront shaping with digital micromirror devices \cite{dremeau2015reference} via beam synthesis with phased-arrays in the microwave domain \cite{lebret1997antenna,fuchs2017antenna} to infrared metamaterial phase holograms \cite{larouche2012infrared}. 

Note that we have an analytical forward model and borrow efficient error back-propagation tools (developed for neural network training) to solve a \textit{constrained} inverse problem --- in contrast to other recent efforts to solve \textit{continuous} inverse problems by first training an ANN to approximate a forward model \cite{peurifoy2018nanophotonic,liu2018training}. 
In passing, we thus introduce a simple constrained inverse configuration-design paradigm for dynamic metasurfaces and show how it enables beam synthesis or the radiation of orthogonal patterns, as opposed to the (thus far) conventionally use of random patterns \cite{sleasman2015dynamic,imani2018two}. While the hardware is the same, the identification of appropriate metasurface configurations is an additional one-off effort. With a modified cost-function, one can also identify settings for scene illuminations with custom-tailored speckle statistics, which may drastically improve the efficiency of computational microwave ghost imaging \cite{bender2018customizing,diebold2018phaseless,kuplicki2016high}.

\subsection{Taxonomy of Illumination Strategies in Wave-Based Sensing}

\begin{figure}[h]
	\begin{center}
\includegraphics [width=\columnwidth] {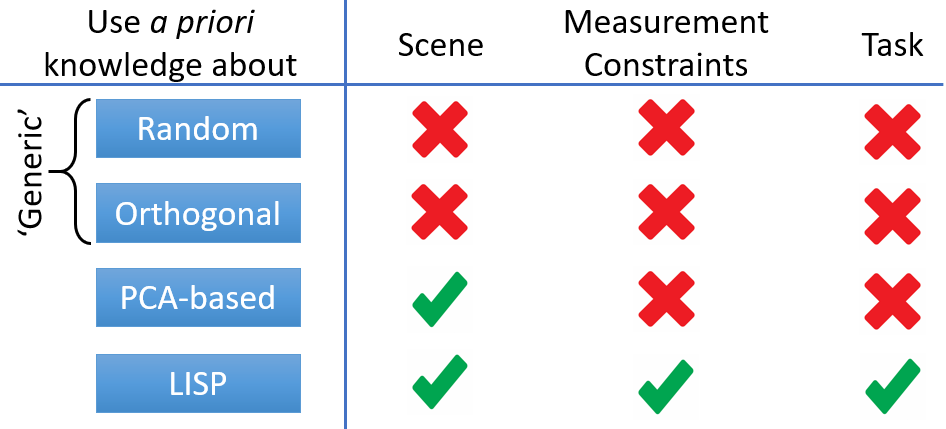}
	\caption{Taxonomy of illumination strategies in wave-based sensing in terms of the \textit{a priori} knowledge they make use of. Schemes using generic (random or orthogonal) scene illuminations ignore any available \textit{a priori} knowledge. Illuminations based on a principle component analysis (PCA) of the scene include knowledge about the scene but not about the hardware constraints or task to be performed. Our proposed Learned Integrated Sensing Pipeline (LISP) makes use of all available knowledge by integrating measurement and processing into a unique ML pipeline. }
	\label{taxonomy}
	\end{center}
\end{figure}

\newpage

\providecommand{\noopsort}[1]{}\providecommand{\singleletter}[1]{#1}%

\end{document}